\documentclass[pra,twocolumn,superscriptaddress]{revtex4} 
\usepackage{epsfig}

\usepackage{graphicx}

\newcommand{\tpaper}{this paper} 

\newcommand{\beq}{\begin{equation}}
\newcommand{\eeq}{\end{equation}}

\newcommand{\commentout}[1]{}

\newcommand{\hide}[1]{}

\newcommand{\eq}[1]{Eq.\,(\ref{#1})}

\newcommand{\fig}[1]{Fig.\,\ref{#1}}

\begin{document}
\title{Fast mode of rotating atoms in one-dimensional lattice rings}
\author{T. Wang}
\affiliation{Department of Physics, University of Connecticut,
Storrs, CT 06269}

\author{S. F. Yelin}
\affiliation{Department of Physics, University of Connecticut,
Storrs, CT 06269} \affiliation{ITAMP, Harvard-Smithsonian Center
for Astrophysics, Cambridge, MA 02138}

\date{\today}
\begin{abstract}
We study the rotation of atoms in one-dimensional lattice rings.
In particular, the ``fast mode", where the ground state atoms
rotate faster than the stirring rotating the atoms, is studied
both analytically and numerically. The conditions for the
transition to the fast mode are found to be very different from
that in continuum rings. We argue that these transition
frequencies remain unchanged for bosonic condensates described in
a mean field. We show that Fermionic interaction and filling
factor have a significant effect on the transition to the fast
mode, and Pauli principle may suppress it altogether.
\end{abstract}

\maketitle

Recent progress in manipulating neutral atoms includes paired
states of fermions,  the crossover between a Bose-Einstein
condensate (BEC) molecules and a Bardeen-Cooper-Schrieffer (BCS)
superfluid of loosely bound pairs
~\cite{Jin03,FermionCondensation,MolecularPop,DressedMolecules,Ohashi,VorticesBECBCS,RotFermi}.
So far, most of the experimental studies involving cold atoms were
conducted in continuum. Optical lattices, however, allow many
effects of interest: Superfluid to Mott insulator
transition~\cite{MottGreiner}, Bloch oscillation of particles in
lattices due to Bragg scattering~\cite{Kittel}, parametric atomic
down conversion in BEC ~\cite{ParametricLatt}, etc.

By rotating atoms in a continuum, many phenomena have already been
investigated: the appearance of
vortices~\cite{VortexNucleation,FastRotBEC} in both the BEC and
BCS~\cite{VorticesBECBCS}, quantum Hall states for fermions with
fast rotation frequency~\cite{HoRotFermi} and vortex lattices in
the lowest Landau level for
BECs~\cite{FastRotBEC,VortexLatticeLLL}. Experimentally, these
rotations are realized by stirring the cold atoms. When the atoms
are rotated in lattices, theoretical studies show that lattices
lead to many new effects under rotation, such as structural phase
transitions of vortex matter~\cite{StructuralPhaseTransOL}. Also,
near the superfluid--Mott insulator transition, the vortex core
has a tendency toward the Mott insulating
phase~\cite{WuVortexconfigurations} and second-order quantum phase
transitions between states of different symmetries in a two
dimensional (2D) lattice were observed at discrete rotation
frequencies~\cite{CarrRot,Bhat06}. In particular, we have recently
shown~\cite{rorHubbard2D} that it is possible for atoms to stay in
a 2D lattice even if the rotation frequency is larger than the
harmonic trapping frequency and density depletion in the trap can
then be developed in such regime.

However, in a 2D lattice, analytic results are not easy to obtain.
The origin of some phenomena is thus hard to catch. The so-called
fast mode for rotating atoms is such an example. This mode was
shown to exist in 2D lattices~\cite{CarrRot,Bhat06}, but we
investigate this here in a 1D lattice through both analysis and
numerical simulations. Throughout this article, we define a
current to be ``positive" if it is faster than the stirring force
and ``negative" if it is slower. Thus, the ground state (GS) is in
the ``fast mode" if its current is positive. While the ring
lattice is much simpler, it catches much of the physics of the
fast mode in higher dimensions, too. Specifically, we will study
the origin of the fast mode, its dependence on quantum statistics
and whether the fast mode is always present when the rotation
frequency is large enough. Before we move on, we remark that
stirring a classical fluid can never achieve positive current in
the rotating frame and  thus the fast mode does not exist for a
classical fluid.

For completeness, we first review the rotation of a single quantum
particle in a
continuum ring with radius $R$. 
In the rotating frame, the Hamiltonian for such system is \mbox{
$H=-\frac{\hbar^2}{2mR^2}\frac{\partial^2}{\partial \phi^2}-\Omega
L_z$}, where $L_z$ is the angular momentum operator
$-i\hbar\partial/\partial \phi$, $m$ is the mass, $\phi$ is the
angular coordinate and $R$ is the radius of the ring. $H$ can be
reduced to $H=-\frac{\hbar^2}{2mR^2}(\frac{\partial}{\partial
\phi}-i\frac{m\Omega R^2}{\hbar})^2-m\Omega^2R^2/2$ with
eigenfunctions $\psi_n(\phi)=\exp{(in\phi)}/\sqrt{2\pi}$, where
$n$ are integers. Except $m\Omega^2R^2/2$ in $H$ which obviously
reflects the centrifugal potential, the Hamiltonian is identical
to that of a $q$-charged particle put into a ring thread by a flux
$\phi=2\pi m\Omega R^2c/q$~\cite{FluxThroughRing}. 
The spectrum $E_n=\frac{\hbar^2}{2mR^2}(n-\frac{m\Omega
R^2}{\hbar})^2-m\Omega^2R^2/2$ of the Hamiltonian shows that when
$m\Omega R^2/\hbar$ is a half integer $k+1/2$, all energy levels
become two-fold degenerate, $E_{k+l+1}=E_{k-l}$; For all other
values of $\Omega$, the ground states are not degenerate.
Furthermore, when $m\Omega R^2/\hbar$ is a half integer, the
doubly degenerate ground states ($n=0,1$) can have either positive
particle currents (as is the case for $n=1$) or negative particle
current ($n=0$) in the rotating frame. For all other cases, the
non-degenerate ground states ($n=0$) have negative currents in the
rotating frame. In what follows, we will show that lattices change
many of these results.

We now study the rotation of bosonic and fermionic atoms in a
lattice ring using a Hubbard model. In the rotating frame, the
single band Boson Hubbard Hamiltonian
is~\cite{Zoller98,CarrRot,Bhat06}
\begin{equation}
\label{eq:LatticeH} H  =  (\sum_{\langle
i,j\rangle}\left(-t-i\Omega K_{i,j}\right)b^+_ib_j+H.c.)+U_B
\end{equation}
where $b^+_i$ ($b_i$) is the Boson creation (annihilation)
operator at site $i$, and the modified single band Fermion Hubbard
Hamiltonian is~\cite{Zoller98,CarrRot,Bhat06}

\begin{equation}
\label{eq:LatticeHF} H  =  (\sum_{\langle
i,j\rangle,\sigma}\left(-t-i\Omega
K_{i,j}\right)c^+_{i,\sigma}c_{j,\sigma}+H.c.)+U_F
\end{equation}
where $c^+_{i,\sigma}$ ($c_{i,\sigma}$) is the  creation
(annihilation) operator for a fermion of spin $\sigma=\uparrow,
\downarrow$ at site $i$. In the
Eqs.~(\ref{eq:LatticeH}-\ref{eq:LatticeHF}), $\langle i,j\rangle$
indicates nearest neighbor pairs, $t$ is the hopping term between
neighboring sites, and $H.c.$ means Hermitian conjugate. The
interaction term for bosons (fermions) is
$U_B=U\sum_i(b^+_ib_i-1)b^+_ib_i$
($U_F=U\sum_ic^+_{i,\uparrow}c_{i,\uparrow}c^+_{i,\downarrow}c_{i,\downarrow}$)
with $U$ the interaction strength determined by the s-wave
scattering length and the lattice potential. 
With equally spaced sites as exemplified by this paper, the
geometric factor $K_{i,j}$ is given by $K=\beta\sin\alpha/2$,
where $\alpha$ is the angle subtended by neighboring sites with
respect to the axis of rotation, and $\beta$ is a dimensionless
constant of order $1$ characterizing the lattice geometry and
depth~\cite{CarrRot,Bhat06,rorHubbard2D}. In
Eqs.~(\ref{eq:LatticeH},\ref{eq:LatticeHF}), the lattice constant
and $\hbar$ are set to be one, so energies are given in units of
the hopping energy. Note that the Hamiltonian used in
Eqs.~(\ref{eq:LatticeH},\ref{eq:LatticeHF}) is an approximation
based on a perturbative treatment of $\Omega L_z$: The Wannier
basis states are the eigenstates of the Hamiltonian without this
term. While $\Omega L_z$ may be included for an alternative (and
possibly better) Wannier basis, Ref.~\cite{Bhat06} (e.g., Fig. 3
therein) shows that the differences are negligible if $\Omega$ is
small. This, however, is the regime we are interested in here.

As a building block for the following study, a single atom in the
lattice with $N_A$ sites is studied first. Since wavefunctions
have to be unique, the total winding phase $\Theta$ around the
ring has to be a multiple of $2\pi$, e.g., $\Theta(n)=n2\pi$ with
n a nonnegative integer determining the symmetry of the state. The
wavefunctions in an evenly distributed lattice site system are
totally determined by the phase $\phi(n)=\Theta(n)/N_A$ as
$|\psi\rangle=\sum_j\exp{\left(i\,\phi\,
j\right)}/\sqrt{N_A}\,\,|j\rangle$, where $|j\rangle$ is the state
when the particle is at site $j$. We can then use the
wavefunctions to determine their energies, currents (see
Fig.~\ref{fig:allCR8}) etc.. In particular, the energies for the
low lying states are $E(n)=-2\left(t\cos{\phi(n)}+\Omega
K\sin{\phi(n)}\right)$, a linear function of rotation frequency
$\Omega$ for any fixed n. From $E(n)$, we see that for $K\Omega\gg
t$, the ground state has $n=n_m\equiv\lfloor N_A/4\rfloor$, where
$\lfloor x\rfloor$ is the the largest integer number that is less
than or equal to the specified number $x$. This means that any
state with maximum phase differences of $\pi/2$ between
neighboring sites are ground state for large $\Omega$.
Furthermore, the linear dependence of $E(n)$ on $\Omega$ should be
contrasted to the quadratic dependence in continuum rings.  Since
the discrete rotational symmetry is not broken by the rotation
(the $\Omega$ term in
Eqs.~{(\ref{eq:LatticeH}-\ref{eq:LatticeHF})), states with
different symmetry experience level crossings as $\Omega$ changes.
By comparing the energies of states with different $n$, the level
crossings between these states can be determined analytically.
Figure~\ref{fig:allCR8}a is an example of the energies and the
level crossings for the low lying states of an atom in an  8-site
ring lattice. These level crossings demonstrate how the system
evolves between states of different rotational symmetries as
$\Omega$ increases~\cite{CarrRot,Bhat06}.

\begin{figure}[ht]
\centerline{\includegraphics[clip,width=.8\linewidth]{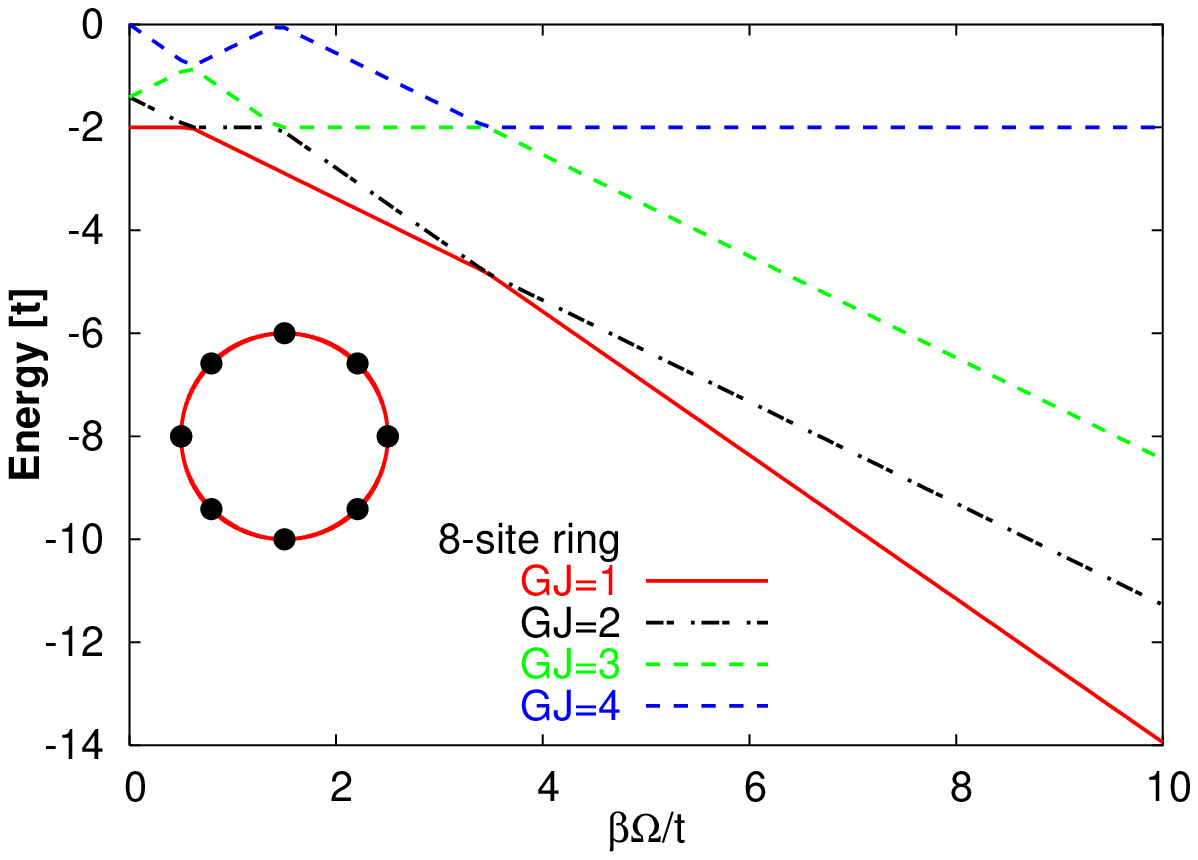}}
\centerline{\includegraphics[clip,width=.8\linewidth]{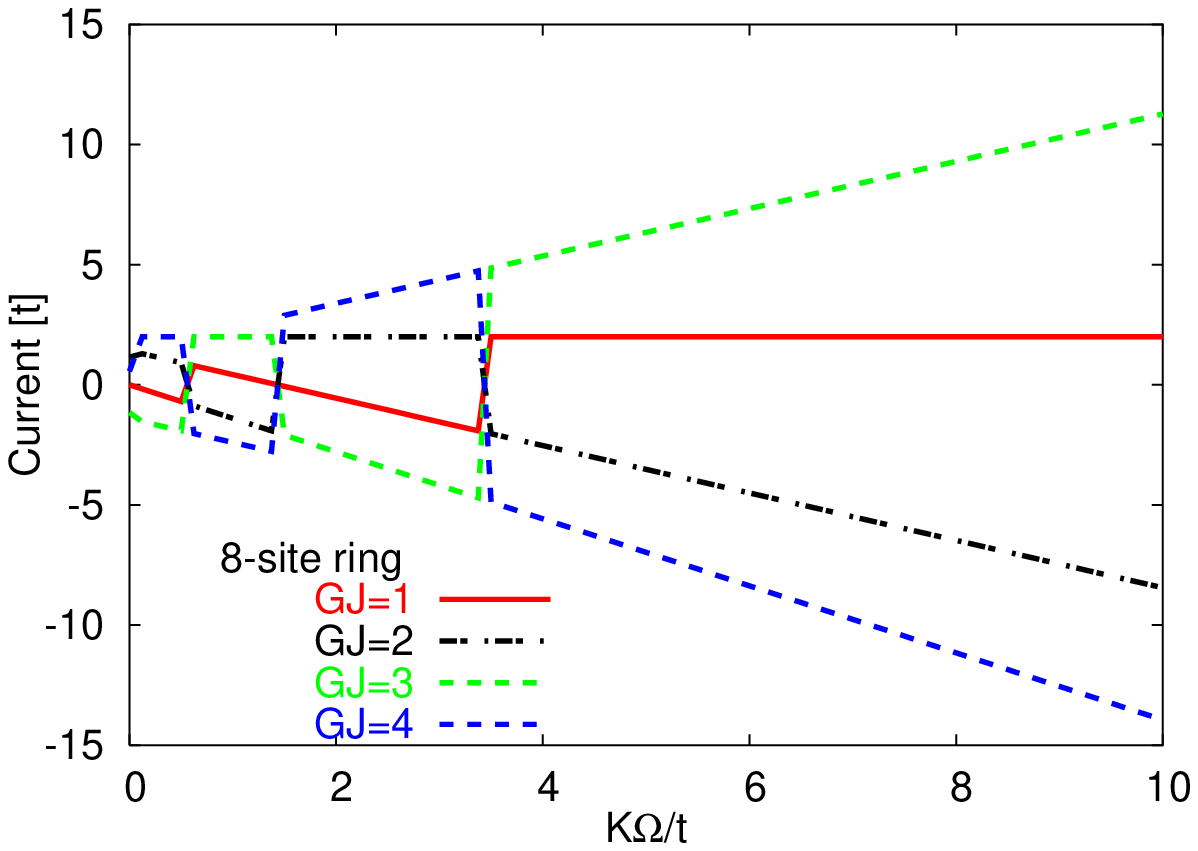}}
    \caption{\protect\label{fig:allCR8} Analytically calculated energies (a)
currents (b) of the low lying states of an atom in the ring
lattice. The inset in (a) shows a sketch of a rotating ring
lattice with eight evenly distributed sites. ``GJ" means the index
of the low lying states. For example, GJ=1 is for the ground state
and GJ=2 is for the first excited state, and so on. At
$\Omega>\Omega_c$, the GS current is a constant $2t$, because the
phase difference $\pi/2$ between neighboring-sites makes $\Omega$
dependent term in Eq.~(\ref{eq:Current}) have no contribution to
the current. }
\end{figure}

Corresponding to theses level crossings, the current and other
observables change as well. As an important observable, the
current is used in what follows to study the fast mode in the
lattice ring. The currents are shown in
Figure~\ref{fig:allCR8}b, where the particle current between site i and j is calculated by~\cite{CarrRot,Bhat06} 
\begin{equation}
J_{ij}=i[n_i,H_{ij}]=it(a_ia^+_j-H.c.)+ K\Omega(a_ia^+_j+H.c.)
\label{eq:Current}
\end{equation}
for each individual state. All the currents $J$ plotted in
\tpaper\ are the integrated currents along the ring. In a
homogeneous ring, they are given by $J=N_s\,J_{ij}$. It is not
surprising to see the persistent currents of the excited states
can be bigger than that of GS or have opposite direction. The
excited states will be used later to construct the state of spin
polarized fermions.

The fast mode in the lattice ring happens when the winding number
of GS is at its maximum $n_m$, where neighboring sites have
maximum phase difference $n_m2\pi/N_A$ and thus maximum currents
are achieved. By matching the energy of the state with winding
number $(n_m-1)$ with the energy of the state with winding number
$n_m$, we obtain $\Omega_c K (1-\cos{2\pi/N_A})=t\sin{2\pi/N_A}$
to
determine the rotational frequency $\Omega_c$ above which GS is in the fast mode. 
Setting $N_A=4$ reproduces the special result in
Ref.~\cite{CarrRot,Bhat06}. From our general result, it can be
seen that for larger $N_A$, $\Omega_c$ is higher to achieve the
fast mode, which makes it difficult to be realized experimentally.
In addition, a constant current of $2t$ is achieved when the phase
difference of $\pi/2$ between neighboring sites makes the current
independent of $\Omega$ (see \eq{eq:Current} and
\fig{fig:allCR8}). For comparison, particles in continuum rings
can move faster than the stirring only when $m\Omega R^2/\hbar$ is
a half integer. The difference is because the lattice breaks the
continuous rotational symmetry.

A single atom in a few-site ring does not answer the question of
whether the fast mode comes from finite number of atoms or finite
number of sites. Thus we study the same ring but with more than
one atom, in which case the quantum statistics of the atoms comes
into play. First we consider a BEC consisting of many
zero-temperature bosons described by a mean field. When the ring
is homogeneous, the distribution of the atoms in the lattice does
not depend on the rotation. This means that the mean field
wavefunction is the same as that of the single-atom wavefunction
except normalized to $N$ atoms, and thus the transition to fast
mode does not change. Since for a BEC many bosons are involved,
the fast mode is therefore only due to finite number of sites in
the lattice ring.

Spin polarized fermions at zero temperature show more interesting
dynamics. The GS current of N spin polarized fermions can be
obtained by summing over the currents of the lowest N states of a
single particle in the lattice, because these fermions, assuming
no p-wave interaction, occupy the lowest N states according to
Pauli's principle. The resulting current is a linear function of
rotation frequency and, at large $\Omega$ can be either positive
(eg. with 3 fermions) or negative (eg. with 2 fermions) as can be
checked from Fig.~\ref{fig:allCR8}b. Therefore, the fast mode may
disappear even when there are only two spin polarized fermions.
This disappearance of the fast mode is due to the Pauli principle,
because Pauli principle precludes occupying the same quantum
state, including the ground state, by identical fermions. This
will be further discussed in the following.

When the fermions are not spin polarized, there can be s-wave
interactions. The simplest example to include this interaction
(see Eq.~\ref{eq:LatticeHF}) is to have one spin up fermion and
one spin down fermion in the ring lattice.
Fig.~\ref{fig:CurrentROmega}a shows the current per fermion as a
function of rotation frequency at various interactions for the two
fermions. Similar to the single atom case, the current approaches
constant $2t$  as the rotation frequency $\Omega$ increases.
However, this asymptotic process depends on the interaction and
can be very slow. In general, the current is not necessarily  a
linear function of rotation frequency. It is important to note
that because of the increased dimension of the Hilbert space,
there are more level crossings than that for single atom as shown
in Fig.~\ref{fig:CurrentROmega}. Although there maybe accidental
degeneracies, generally the rotation frequencies at which level
crossings happen do not coincide for different $U$. At large
rotation frequency (shown in Fig.~\ref{fig:CurrentROmega}b), the
current is always positive, that is, the current changes
continuously from noninteracting case to strongly interacting
regime, to be compared with a system with more fermions discussed
later in this paper. With even larger $\Omega$, the current will
approach $2t$ for all $U$ in this non-spin-polarized two-atom
system.

\begin{figure}[ht]
    \centerline{\includegraphics[clip,width=.8\linewidth]{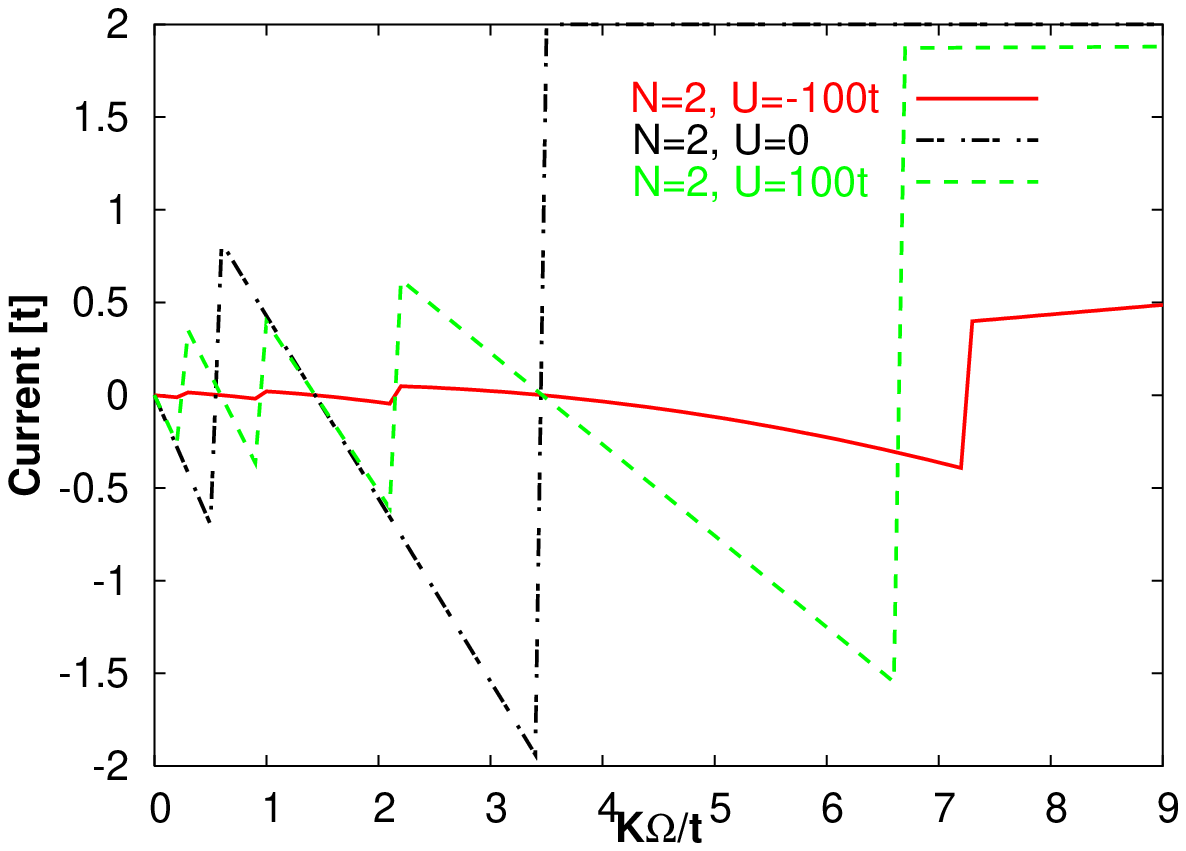}}
\centerline{\includegraphics[clip,width=.8\linewidth]{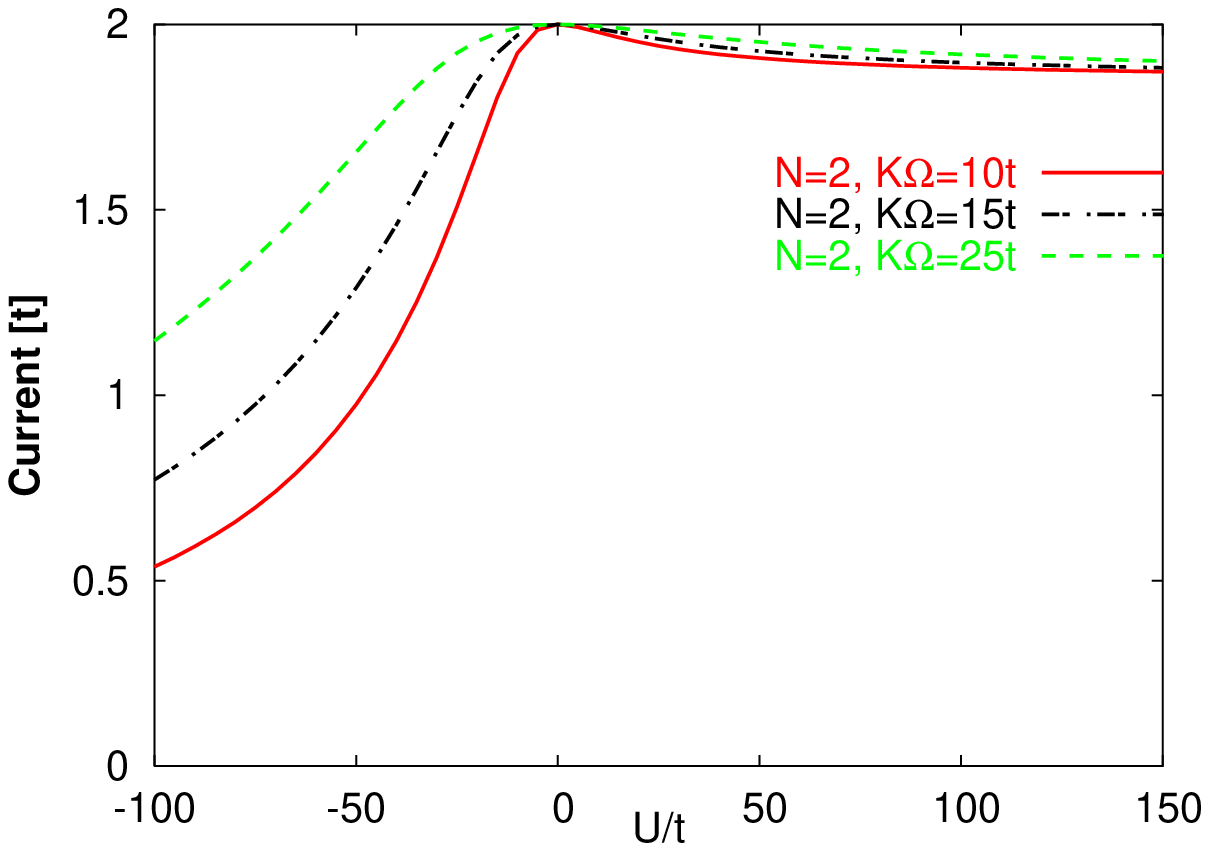}}
    \caption{\protect\label{fig:CurrentROmega} Numerically calculated current per fermion
when one spin up fermion and one spin down fermion are rotated in
an eight site-ring lattice. The current per atom always approaches
$2t$ when the rotation frequency is large enough.}
\end{figure}

It is interesting to study combined effects of spin and
interaction by putting four fermions with two up spins and two
down spins in the lattice. Figure~\ref{fig:CurrentRlargeU} shows
the current as a function of the interaction at large rotation
frequencies $\Omega$. With both strong attractive and repulsive
interactions, positive currents and thus the fast mode persist.
However, when the interaction strength is small $U\sim0$, the
currents surprisingly become negative and are nearly linear in the
rotation frequency. The case without interaction, $U=0$, is easy
to understand, because the current is the sum over the currents of
the lowest two states for spin up and spin down fermions shown in
Fig.~\ref{fig:allCR8}b. From the discussion on the spin polarized
fermions, it is clear that the total current for the four
noninteracting fermions is negative (thus not in fast mode) when
$\Omega$ is very large. Furthermore, Fig.~\ref{fig:CurrentRlargeU}
also shows that the absence of fast mode for non-spin polarized
fermions extends to finite attractive and repulsive interactions.
This means that the Pauli principle may suppress the fast mode in
many-fermion system even at nonzero interaction.

\begin{figure}[ht]
\centerline{\includegraphics[clip,width=.8\linewidth]{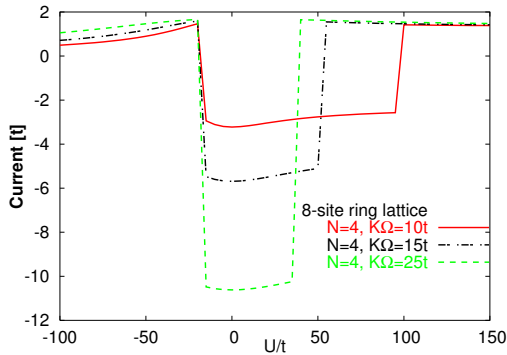}} 
    \caption{\protect\label{fig:CurrentRlargeU}
Fast rotation of four fermions in 8 site-ring lattice. It shows
numerically the effects of interaction and spin on the quantum
phase transition between fast mode and non-fast mode. The non-fast
mode around $U\sim0$ is caused by Fermi pressure (see text). }
\end{figure}

Next, we try to explain the fast mode in such four-fermion system.
At large attractive interaction ($U\ll0$), the fermions form local
pairs~\cite{BoseFermiMix}, and thus behave like bosons, so the
current is positive which is not surprising given that fast
rotating bosons always display a fast mode. Since the fermions get
paired roughly independent of the rotation frequency, the
transition from the fast mode to the non-fast mode happens at
roughly the same $U$, independent of the rotating frequency as
shown in Fig.~\ref{fig:CurrentRlargeU}. While with large repulsive
interaction ($U\gg0$), no such pairing exists and whether the
transition happens at $U$ depends on the rotation frequency. The
larger $\Omega$ is, the smaller repulsive U is required to make
the transition happen (see Fig.~\ref{fig:CurrentRlargeU}).
Furthermore, since the fermions are not paired for strong
repulsive interaction, the existence of fast modes at large
rotation frequency means that the fast mode does not necessarily
mean pairing or condensation of fermions.

Although our study is meant to be an in-principle theoretical
discussion, a few comments about experiments are in order. The
realization of a ring lattice with a tunable boundary phase twist
has been proposed recently~\cite{RealizeRing}. Since many bosons
can be in the same fast mode, time of flight after turning off the
potential of even a single lattice ring should allow detection of
the momentum distribution, and thus the fast mode, of the atoms.
Detecting the fast mode of fermions may need additional twists,
such as using multiple lattice rings, to enhance the signals.
Using multiple lattices to enhance the signal has been
demonstrated recently~\cite{DoubleWellLattice}.

To conclude, we have studied the rotation of bosonic and fermionic
atoms in one dimensional lattice rings. We found that minimizing
the ground state energy may give the fast mode and the transition
to the fast mode in lattice rings is very different from that in
continuum rings. Fermionic interaction and the filling factor are
shown to have significant effects on the transitions to the fast
mode and  Fermi pressure may suppress the fast mode. Finally, the
fast mode is due to the finite number of lattice sites and is not
associated to bosonic statistics, pairing of fermions, or
superfluidity.

This work is supported by NSF and Research Corporation. T. Wang
acknowledges the helpful discussions with J. Javanainen and U.
Shrestha.

\bibliography{BECBCS}

\begin{thebibliography}{24}
\expandafter\ifx\csname natexlab\endcsname\relax\def\natexlab#1{#1}\fi
\expandafter\ifx\csname bibnamefont\endcsname\relax
  \def\bibnamefont#1{#1}\fi
\expandafter\ifx\csname bibfnamefont\endcsname\relax
  \def\bibfnamefont#1{#1}\fi
\expandafter\ifx\csname citenamefont\endcsname\relax
  \def\citenamefont#1{#1}\fi
\expandafter\ifx\csname url\endcsname\relax
  \def\url#1{\texttt{#1}}\fi
\expandafter\ifx\csname urlprefix\endcsname\relax\def\urlprefix{URL }\fi
\providecommand{\bibinfo}[2]{#2}
\providecommand{\eprint}[2][]{\url{#2}}

\bibitem[{\citenamefont{Regal et~al.}(2003)\citenamefont{Regal, Ticknor, Bohn,
  and Jin}}]{Jin03}
\bibinfo{author}{\bibfnamefont{C.~A.} \bibnamefont{Regal}},
  \bibinfo{author}{\bibfnamefont{C.}~\bibnamefont{Ticknor}},
  \bibinfo{author}{\bibfnamefont{J.~L.} \bibnamefont{Bohn}}, \bibnamefont{and}
  \bibinfo{author}{\bibfnamefont{D.~S.} \bibnamefont{Jin}},
  \bibinfo{journal}{Nature} \textbf{\bibinfo{volume}{424}}, \bibinfo{pages}{47}
  (\bibinfo{year}{2003}).

\bibitem[{\citenamefont{Zwierlein et~al.}(2004)\citenamefont{Zwierlein, Stan,
  Schunck, Raupach, Kerman, and Ketterle}}]{FermionCondensation}
\bibinfo{author}{\bibfnamefont{M.~W.} \bibnamefont{Zwierlein}},
  \bibinfo{author}{\bibfnamefont{C.~A.} \bibnamefont{Stan}},
  \bibinfo{author}{\bibfnamefont{C.~H.} \bibnamefont{Schunck}},
  \bibinfo{author}{\bibfnamefont{S.~M.~F.} \bibnamefont{Raupach}},
  \bibinfo{author}{\bibfnamefont{A.~J.} \bibnamefont{Kerman}},
  \bibnamefont{and} \bibinfo{author}{\bibfnamefont{W.}~\bibnamefont{Ketterle}},
  \bibinfo{journal}{Phys.~Rev.~Lett.} \textbf{\bibinfo{volume}{92}},
  \bibinfo{pages}{120403} (\bibinfo{year}{2004}).

\bibitem[{\citenamefont{Chen and Levin}(2005)}]{MolecularPop}
\bibinfo{author}{\bibfnamefont{Q.}~\bibnamefont{Chen}} \bibnamefont{and}
  \bibinfo{author}{\bibfnamefont{K.}~\bibnamefont{Levin}},
  \bibinfo{journal}{Phys.~Rev.~Lett.} \textbf{\bibinfo{volume}{95}},
  \bibinfo{pages}{260406} (\bibinfo{year}{2005}).

\bibitem[{\citenamefont{Romans and Stoof}(2005)}]{DressedMolecules}
\bibinfo{author}{\bibfnamefont{M.~W.~J.} \bibnamefont{Romans}}
  \bibnamefont{and} \bibinfo{author}{\bibfnamefont{H.~T.~C.}
  \bibnamefont{Stoof}}, \bibinfo{journal}{Phys.~Rev.~Lett.}
  \textbf{\bibinfo{volume}{95}}, \bibinfo{pages}{260407}
  (\bibinfo{year}{2005}).

\bibitem[{\citenamefont{Ohashi and Griffin}(2002)}]{Ohashi}
\bibinfo{author}{\bibfnamefont{Y.}~\bibnamefont{Ohashi}} \bibnamefont{and}
  \bibinfo{author}{\bibfnamefont{A.}~\bibnamefont{Griffin}},
  \bibinfo{journal}{Phys.~Rev.~Lett.} \textbf{\bibinfo{volume}{89}},
  \bibinfo{pages}{130402} (\bibinfo{year}{2002}).

\bibitem[{\citenamefont{Zwierlein et~al.}(2005)\citenamefont{Zwierlein,
  Abo-Shaeer, , Schirotzek, Schunck, and Ketterle}}]{VorticesBECBCS}
\bibinfo{author}{\bibfnamefont{M.~W.} \bibnamefont{Zwierlein}},
  \bibinfo{author}{\bibfnamefont{J.~R.} \bibnamefont{Abo-Shaeer}}, ,
  \bibinfo{author}{\bibfnamefont{A.}~\bibnamefont{Schirotzek}},
  \bibinfo{author}{\bibfnamefont{C.~H.} \bibnamefont{Schunck}},
  \bibnamefont{and} \bibinfo{author}{\bibfnamefont{W.}~\bibnamefont{Ketterle}},
  \bibinfo{journal}{Nature} \textbf{\bibinfo{volume}{435}},
  \bibinfo{pages}{1047} (\bibinfo{year}{2005}).

\bibitem[{\citenamefont{Schunck et~al.}(2006)\citenamefont{Schunck, Zwierlein,
  Schirotzek, and Ketterle}}]{RotFermi}
\bibinfo{author}{\bibfnamefont{C.}~\bibnamefont{Schunck}},
  \bibinfo{author}{\bibfnamefont{M.}~\bibnamefont{Zwierlein}},
  \bibinfo{author}{\bibfnamefont{A.}~\bibnamefont{Schirotzek}},
  \bibnamefont{and} \bibinfo{author}{\bibfnamefont{W.}~\bibnamefont{Ketterle}}
  (\bibinfo{year}{2006}), \bibinfo{note}{cond-mat/0607298}.

\bibitem[{\citenamefont{Greiner et~al.}(2002)\citenamefont{Greiner, Mandel,
  Esslinger, HaÈnsch, and Bloch}}]{MottGreiner}
\bibinfo{author}{\bibfnamefont{M.}~\bibnamefont{Greiner}},
  \bibinfo{author}{\bibfnamefont{O.}~\bibnamefont{Mandel}},
  \bibinfo{author}{\bibfnamefont{T.}~\bibnamefont{Esslinger}},
  \bibinfo{author}{\bibfnamefont{T.~W.} \bibnamefont{HaÈnsch}},
  \bibnamefont{and} \bibinfo{author}{\bibfnamefont{I.}~\bibnamefont{Bloch}},
  \bibinfo{journal}{Nature} \textbf{\bibinfo{volume}{415}}, \bibinfo{pages}{39}
  (\bibinfo{year}{2002}).

\bibitem[{\citenamefont{Kittel}(1996)}]{Kittel}
\bibinfo{author}{\bibfnamefont{C.}~\bibnamefont{Kittel}},
  \emph{\bibinfo{title}{Introduction to solid state physics}}
  (\bibinfo{publisher}{John Wiley \& Sons, Inc., New York},
  \bibinfo{year}{1996}), ISBN \bibinfo{isbn}{0-471-11181-3}, \bibinfo{note}{7th
  ed.}

\bibitem[{\citenamefont{Campbell et~al.}(2006)\citenamefont{Campbell, Mun,
  Boyd, Streed, Ketterle, and Pritchard}}]{ParametricLatt}
\bibinfo{author}{\bibfnamefont{G.~K.} \bibnamefont{Campbell}},
  \bibinfo{author}{\bibfnamefont{J.}~\bibnamefont{Mun}},
  \bibinfo{author}{\bibfnamefont{M.}~\bibnamefont{Boyd}},
  \bibinfo{author}{\bibfnamefont{E.~W.} \bibnamefont{Streed}},
  \bibinfo{author}{\bibfnamefont{W.}~\bibnamefont{Ketterle}}, \bibnamefont{and}
  \bibinfo{author}{\bibfnamefont{D.~E.} \bibnamefont{Pritchard}},
  \bibinfo{journal}{Phys. Rev. Lett.} \textbf{\bibinfo{volume}{96}},
  \bibinfo{pages}{020406} (\bibinfo{year}{2006}).

\bibitem[{\citenamefont{Raman et~al.}(2001)\citenamefont{Raman, Abo-Shaeer,
  Vogels, Xu, and Ketterle}}]{VortexNucleation}
\bibinfo{author}{\bibfnamefont{C.}~\bibnamefont{Raman}},
  \bibinfo{author}{\bibfnamefont{J.~R.} \bibnamefont{Abo-Shaeer}},
  \bibinfo{author}{\bibfnamefont{J.~M.} \bibnamefont{Vogels}},
  \bibinfo{author}{\bibfnamefont{K.}~\bibnamefont{Xu}}, \bibnamefont{and}
  \bibinfo{author}{\bibfnamefont{W.}~\bibnamefont{Ketterle}},
  \bibinfo{journal}{Phys. Rev. Lett.} \textbf{\bibinfo{volume}{87}},
  \bibinfo{pages}{210402} (\bibinfo{year}{2001}).

\bibitem[{\citenamefont{Aftalion et~al.}(2005)\citenamefont{Aftalion, Blanc,
  and Dalibard}}]{FastRotBEC}
\bibinfo{author}{\bibfnamefont{A.}~\bibnamefont{Aftalion}},
  \bibinfo{author}{\bibfnamefont{X.}~\bibnamefont{Blanc}}, \bibnamefont{and}
  \bibinfo{author}{\bibfnamefont{J.}~\bibnamefont{Dalibard}},
  \bibinfo{journal}{Phys. Rev. A} \textbf{\bibinfo{volume}{71}},
  \bibinfo{pages}{023611} (\bibinfo{year}{2005}).

\bibitem[{\citenamefont{Ho and Ciobanu}(2000)}]{HoRotFermi}
\bibinfo{author}{\bibfnamefont{T.-L.} \bibnamefont{Ho}} \bibnamefont{and}
  \bibinfo{author}{\bibfnamefont{C.~V.} \bibnamefont{Ciobanu}},
  \bibinfo{journal}{Phys. Rev. Lett.} \textbf{\bibinfo{volume}{85}},
  \bibinfo{pages}{4648} (\bibinfo{year}{2000}).

\bibitem[{\citenamefont{Cooper et~al.}(2004)\citenamefont{Cooper, Komineas, and
  Read}}]{VortexLatticeLLL}
\bibinfo{author}{\bibfnamefont{N.~R.} \bibnamefont{Cooper}},
  \bibinfo{author}{\bibfnamefont{S.}~\bibnamefont{Komineas}}, \bibnamefont{and}
  \bibinfo{author}{\bibfnamefont{N.}~\bibnamefont{Read}},
  \bibinfo{journal}{Phys. Rev. A} \textbf{\bibinfo{volume}{70}},
  \bibinfo{pages}{033604} (\bibinfo{year}{2004}).

\bibitem[{\citenamefont{Pu et~al.}(2005)\citenamefont{Pu, Baksmaty, Yi, and
  Bigelow}}]{StructuralPhaseTransOL}
\bibinfo{author}{\bibfnamefont{H.}~\bibnamefont{Pu}},
  \bibinfo{author}{\bibfnamefont{L.~O.} \bibnamefont{Baksmaty}},
  \bibinfo{author}{\bibfnamefont{S.}~\bibnamefont{Yi}}, \bibnamefont{and}
  \bibinfo{author}{\bibfnamefont{N.~P.} \bibnamefont{Bigelow}},
  \bibinfo{journal}{Phys. Rev. Lett.} \textbf{\bibinfo{volume}{94}},
  \bibinfo{pages}{190401} (\bibinfo{year}{2005}).

\bibitem[{\citenamefont{Wu et~al.}(2004)\citenamefont{Wu, Chen, Hu, and
  Zhang}}]{WuVortexconfigurations}
\bibinfo{author}{\bibfnamefont{C.}~\bibnamefont{Wu}},
  \bibinfo{author}{\bibfnamefont{H.-d.} \bibnamefont{Chen}},
  \bibinfo{author}{\bibfnamefont{J.-p.} \bibnamefont{Hu}}, \bibnamefont{and}
  \bibinfo{author}{\bibfnamefont{S.-C.} \bibnamefont{Zhang}},
  \bibinfo{journal}{Phys. Rev. A} \textbf{\bibinfo{volume}{69}},
  \bibinfo{pages}{043609} (\bibinfo{year}{2004}).

\bibitem[{\citenamefont{Bhat et~al.}(2006{\natexlab{a}})\citenamefont{Bhat,
  Holland, and Carr}}]{CarrRot}
\bibinfo{author}{\bibfnamefont{R.}~\bibnamefont{Bhat}},
  \bibinfo{author}{\bibfnamefont{M.~J.} \bibnamefont{Holland}},
  \bibnamefont{and} \bibinfo{author}{\bibfnamefont{L.~D.} \bibnamefont{Carr}},
  \bibinfo{journal}{Phys. Rev. Lett.} \textbf{\bibinfo{volume}{96}},
  \bibinfo{pages}{060405} (\bibinfo{year}{2006}{\natexlab{a}}).

\bibitem[{\citenamefont{Bhat et~al.}(2006{\natexlab{b}})\citenamefont{Bhat,
  Peden, Seaman, Krämer, Carr, and Holland}}]{Bhat06}
\bibinfo{author}{\bibfnamefont{R.}~\bibnamefont{Bhat}},
  \bibinfo{author}{\bibfnamefont{B.~M.} \bibnamefont{Peden}},
  \bibinfo{author}{\bibfnamefont{B.~T.} \bibnamefont{Seaman}},
  \bibinfo{author}{\bibfnamefont{M.}~\bibnamefont{Krämer}},
  \bibinfo{author}{\bibfnamefont{L.~D.} \bibnamefont{Carr}}, \bibnamefont{and}
  \bibinfo{author}{\bibfnamefont{M.~J.} \bibnamefont{Holland}},
  \bibinfo{journal}{Phys.~Rev.~A} \textbf{\bibinfo{volume}{74}},
  \bibinfo{pages}{063606} (\bibinfo{year}{2006}{\natexlab{b}}).

\bibitem[{\citenamefont{Wang and Yelin}(2006)}]{rorHubbard2D}
\bibinfo{author}{\bibfnamefont{T.}~\bibnamefont{Wang}} \bibnamefont{and}
  \bibinfo{author}{\bibfnamefont{S.~F.} \bibnamefont{Yelin}}
  (\bibinfo{year}{2006}), \bibinfo{note}{quant-ph/0610114}.

\bibitem[{\citenamefont{Merzbacher}(1962)}]{FluxThroughRing}
\bibinfo{author}{\bibfnamefont{E.}~\bibnamefont{Merzbacher}},
  \bibinfo{journal}{Amer. J. Phys.} \textbf{\bibinfo{volume}{30}},
  \bibinfo{pages}{237} (\bibinfo{year}{1962}).

\bibitem[{\citenamefont{Jaksch et~al.}(1998)\citenamefont{Jaksch, Bruder,
  Cirac, Gardiner, and Zoller}}]{Zoller98}
\bibinfo{author}{\bibfnamefont{D.}~\bibnamefont{Jaksch}},
  \bibinfo{author}{\bibfnamefont{C.}~\bibnamefont{Bruder}},
  \bibinfo{author}{\bibfnamefont{J.~I.} \bibnamefont{Cirac}},
  \bibinfo{author}{\bibfnamefont{C.~W.} \bibnamefont{Gardiner}},
  \bibnamefont{and} \bibinfo{author}{\bibfnamefont{P.}~\bibnamefont{Zoller}},
  \bibinfo{journal}{Phys. Rev. Lett.} \textbf{\bibinfo{volume}{81}},
  \bibinfo{pages}{3108} (\bibinfo{year}{1998}).

\bibitem[{\citenamefont{Carr and Holland}(2005)}]{BoseFermiMix}
\bibinfo{author}{\bibfnamefont{L.~D.} \bibnamefont{Carr}} \bibnamefont{and}
  \bibinfo{author}{\bibfnamefont{M.~J.} \bibnamefont{Holland}},
  \bibinfo{journal}{Phys. Rev. A} \textbf{\bibinfo{volume}{72}},
  \bibinfo{pages}{031604} (\bibinfo{year}{2005}).

\bibitem[{\citenamefont{Amico et~al.}(2005)\citenamefont{Amico, Osterloh, and
  Cataliotti}}]{RealizeRing}
\bibinfo{author}{\bibfnamefont{L.}~\bibnamefont{Amico}},
  \bibinfo{author}{\bibfnamefont{A.}~\bibnamefont{Osterloh}}, \bibnamefont{and}
  \bibinfo{author}{\bibfnamefont{F.}~\bibnamefont{Cataliotti}},
  \bibinfo{journal}{Phys.~Rev.~Lett.} \textbf{\bibinfo{volume}{95}},
  \bibinfo{pages}{063201} (\bibinfo{year}{2005}).

\bibitem[{\citenamefont{Sebby-Strabley
  et~al.}(2006)\citenamefont{Sebby-Strabley, Anderlini, Jessen, and
  Porto}}]{DoubleWellLattice}
\bibinfo{author}{\bibfnamefont{J.}~\bibnamefont{Sebby-Strabley}},
  \bibinfo{author}{\bibfnamefont{M.}~\bibnamefont{Anderlini}},
  \bibinfo{author}{\bibfnamefont{P.~S.} \bibnamefont{Jessen}},
  \bibnamefont{and} \bibinfo{author}{\bibfnamefont{J.~V.} \bibnamefont{Porto}},
  \bibinfo{journal}{Phys. Rev. A} \textbf{\bibinfo{volume}{73}},
  \bibinfo{pages}{033605} (\bibinfo{year}{2006}).

\end{thebibliography}


\end{document}